%% file: main.tex
\documentclass[a4paper]{article}

\usepackage{INTERSPEECH2022}
\usepackage{tikz}
\usepackage{comment}
\usepackage{amsmath,amssymb,graphicx} % define this before the line numbering.
\usepackage{tipa}
\usepackage{array,booktabs}
\usepackage{color}
\usepackage{cite}
\usepackage{makecell}
\usepackage{multirow}
\usepackage{multicol}
\usepackage{amsmath,graphicx}
\usepackage{amssymb}
\usepackage{booktabs}
\newcommand{\ie}{\textit{i}.\textit{e}.}
\newcommand{\eg}{\textit{e}.\textit{g}.}
\newcommand{\etal}{\textit{et al}.}
\usepackage{xcolor}
\usepackage{lipsum}
\usepackage{arydshln}

\makeatletter
\def\adl@drawiv#1#2#3{%
        \hskip.5\tabcolsep
        \xleaders#3{#2.5\@tempdimb #1{1}#2.5\@tempdimb}%
                #2\z@ plus1fil minus1fil\relax
        \hskip.5\tabcolsep}
\newcommand{\cdashlinelr}[1]{%
  \noalign{\vskip\aboverulesep
           \global\let\@dashdrawstore\adl@draw
           \global\let\adl@draw\adl@drawiv}
  \cdashline{#1}
  \noalign{\global\let\adl@draw\@dashdrawstore
           \vskip\belowrulesep}}
\makeatother

\usepackage[pagebackref,breaklinks,colorlinks,hyperfootnotes=false]{hyperref}

\title{Visual Context-driven Audio Feature Enhancement\\for Robust End-to-End Audio-Visual Speech Recognition}
\name{Joanna Hong$^{1*}$\thanks{$^*$Both authors have contributed equally to this work.}, Minsu Kim$^{1*}$\footnotemark[1], Daehun Yoo$^2$, Yong Man Ro$^{1\dagger}$\thanks{$^\dagger$Corresponding author.}}
\address{
  $^1$KAIST, Daejeon, South Korea\\
  $^2$Genesis Lab Inc., Seoul, South Korea}
\email{joanna2587@kaist.ac.kr, ms.k@kaist.ac.kr, dhyoo@genesislab.ai, ymro@kaist.ac.kr}

\begin{document}

\maketitle
\begin{abstract}
This paper focuses on designing a noise-robust end-to-end Audio-Visual Speech Recognition (AVSR) system. To this end, we propose Visual Context-driven Audio Feature Enhancement module (V-CAFE) to enhance the input noisy audio speech with a help of audio-visual correspondence. The proposed V-CAFE is designed to capture the transition of lip movements, namely visual context and to generate a noise reduction mask by considering the obtained visual context. Through context-dependent modeling, the ambiguity in viseme-to-phoneme mapping can be refined for mask generation. The noisy representations are masked out with the noise reduction mask resulting in enhanced audio features. The enhanced audio features are fused with the visual features and taken to an encoder-decoder model composed of Conformer and Transformer for speech recognition. We show the proposed end-to-end AVSR with the V-CAFE can further improve the noise-robustness of AVSR. The effectiveness of the proposed method is evaluated in noisy speech recognition and overlapped speech recognition experiments using the two largest audio-visual datasets, LRS2 and LRS3.

\end{abstract}
\noindent\textbf{Index Terms}: audio-visual speech recognition, audio enhancement, visual context-driven audio feature enhancement

\section{Introduction}
Using complementary information of different modalities, audio and video, is proved to be effective in modeling speech into both text and acoustic forms \cite{mroueh2015deep, kim2021multi, zhao2020hearing, hong2021speech, ephrat2018looking, kim2021cromm}. Audio-Visual Speech Recognition (AVSR) is one way of employing the complementary effect of audio and visual inputs to improve speech recognizing performance. Since the visual information is not affected by noisy environments such as a crowded restaurant or a construction field, the AVSR system can successfully recognize the corrupted audio speech from those environments by leveraging the visual information.

With its technical demand and the development of deep learning, AVSR techniques have made significant improvements. Initial work \cite{noda2015audio} introduces a connectionist-hidden Markov model with a pretrained CNN to classify phonemes, and recent works further improve the neural architecture and training schemes. Afouras \etal \cite{afouras2018deep} utilize spectrograms for the audio features and the visual features extracted from a pre-trained visual front-end to train a Transformer-based encoder-decoder model \cite{vaswani2017attention}. An end-to-end model for AVSR is introduced through combining the feature extraction and recognition stages by Petridis \etal \cite {petridis2018end}, and Ma \etal \cite{ma2021end} further develop the model architecture using Conformer \cite{gulati2020conformer}.

Even though the AVSR model can enhance the recognition performance under noisy environments by complementing the audio with visual information, explicitly enhancing the noisy audio representations before the recognizing procedure can make the model further robust to the noise. To this end, Xu \etal \cite{xu2020discriminative} propose a two-stage speech recognition model that firstly enhances the speech spectrogram using a mask generated from the visual streams and recognizes the speech using Element-wise-Attention Gated Recurrent Unit. Yu \etal \cite{yu2020overlaplrs2} propose an end-to-end system for audio enhancement and recognition through audio-visual modality driven gated fusion.

For the accurate speech modeling, the context-dependent model \cite{dahl2011context} is utilized to consider the transitions between words which can be called phonetic context. Likewise, Kim \etal \cite{kim2021lip} shows that considering lip appearances back and forth of a given moment is beneficial for refining the viseme-to-phoneme mapping by jointly considering the visual context information. Therefore, by considering the visual context in generating a noise reduction mask from visual information, more accurate speech enhancement for AVSR can be achieved.

In this paper, we try to improve the audio enhancement stage of the AVSR model, which is used to be designed as a simple fusion of audio and visual modalities (\ie, concatenation and passing neural network) \cite{xu2020discriminative,yu2020overlaplrs2}, by proposing Visual Context-driven Audio Feature Enhancement module (V-CAFE). The proposed V-CAFE explicitly incorporates the visual context through cross-modal attention to generate a noise reduction mask. The generated mask is applied to the encoded audio features for reducing the noisy audio representations. Thus, with the visual context, the ambiguity induced by one-to-many mapping of viseme-to-phoneme can be mitigated when enhancing the audio representation. The enhanced audio features from the V-CAFE module are then fused with the visual features to perform the speech recognition. We adopt Conformer-Transformer encoder-decoder architecture of \cite{ma2021end} with the joint CTC/Attention \cite{kim2017joint} loss function. The entire architecture including the V-CAFE is trained in an end-to-end manner. We evaluate the effectiveness of the proposed method in both noisy speech and overlapped speech recognition experiments on LRS2 and LRS3, two largest in-the-wild audio-visual speech datasets. The experimental results verify that the proposed V-CAFE can achieve the robust speech recognition performances under several noisy environments.

%------------------------------------ Figure 1
%#####################################################
\input{./figure/figure1.tex}
%#####################################################

\section{Methodology}
Let $(x_v \in \mathbb{R}^{T \times H \times W \times C}, x_a \in \mathbb{R}^{F \times S}, y \in \mathbb{R}^{L})$ be a pair of lip video, log mel-spectrogram converted from speech audio, and ground-truth label, where $T$ is total frames of video, $H$, $W$, and $C$ are height, width, and channel of a frame, respectively, $F$ and $S$ represent mel-frequency channels and frame length, respectively, and $L$ refers to the length of transcription. Our objective is to develop a robust AVSR model to noise by enhancing the noisy corrupted audio representations with audio-visual correspondence (See Fig.\ref{fig:1}).

\subsection{Visual Context-driven Audio Feature Enhancement}
As phonetic contexts (\eg, triphones) are significant factors in speech recognition \cite{dahl2011context}, it is beneficial to know the front and back lip movements at the given moment in modeling the accurate speech \cite{kim2021lip}. Thus, through the V-CAFE module, we try to get rid of the noises from the input corrupted audio representation by considering not only the current lip appearances but also the visual context obtained from both local and global lip movements. To this end, visual features $f_v$ and audio features $f_a$ are extracted with their own front-end as follows, $f_v=F_v(x_v)\in\mathbb{R}^{T\times D_1}$ and $f_a=F_a(x_a)\in\mathbb{R}^{T\times D_1}$, where $F_v(\cdot)$ and $F_a(\cdot)$ represent visual front-end and audio front-end, respectively. Please note that the frame length of both audio and visual features are designed to be the same. Then, the visual context is acquired at the given time step $t$ through cross-modal attention as follows,
\begin{align}
    C^t=\text{softmax}(\frac{Q^t \cdot K^\top}{\sqrt{D_2}})\cdot V,
\end{align}
\vspace{-0.3cm}
\begin{align}
    Q^t=f_a^t \cdot W_q, \quad K=f_v \cdot W_k, \quad V=f_v \cdot W_v,
\end{align}
where $C^t$ is the obtained visual context for the $t$-th audio feature, $Q^t$, $K$, and $V$ are the embedded query, key, and value for the cross-modal attention, and $W_q\in\mathbb{R}^{D_1\times D_2}$, $W_k\in\mathbb{R}^{D_1\times D_2}$, and $W_v\in\mathbb{R}^{D_1\times D_2}$ represent the embedding weights for the query, key, and the value, respectively. Through the cross-modal attention, both the local and global lip movements are embedded for enhancing the given noisy audio features.

With the visual context $C=\{C^1,\dots,C^T\}\in\mathbb{R}^{T\times D_2}$, we predict the speech-related representations from the audio features $f_a$, which have correlation with the lip movements. To this end, a mask generator is introduced which predicts the mask to be applied to the audio features for suppressing the noise. The mask generator is composed of two convolution layers with ReLU and Sigmoid activation respectively as follows,
\begin{align}
m = \text{Sigmoid}(\text{Conv}(\text{ReLU}(\text{Conv}(C;\theta_1));\theta_2)),
\end{align}
where $m\in\mathbb{R}^{T\times D_1}$ is the generated mask whose values are in range $\left[0, 1\right]$, Conv$(\,\cdot\,;\theta)$ represents the 1D convolution layer with the weight parameter of $\theta$. 

The mask is multiplicated to the audio features $f_a$ so that the noise representations in the audio can be masked out while the meaningful speech representations can remain. The masked audio features are summed with the original features to obtain the enhanced audio features as follows,  
\begin{align}
    \hat{f}_a = (f_a \odot m) + f_a,
\end{align}
where $\hat{f}_a$ is the enhanced audio features by considering the visual context and $\odot$ is element-wise multiplication. With the enhanced audio features $\hat{f}_a$, we can perform noise-robust AVSR. The detailed illustration of the V-CAFE can be found in Fig.\ref{fig:2}.

%------------------------------------ Figure 2
%#####################################################
\input{./figure/figure2.tex}
%#####################################################

\subsection{Noise-Robust End-to-End AVSR via V-CAFE}
The proposed method is trained in an end-to-end manner including the front-ends, V-CAFE, and a sequential encoder-decoder module. We employ an architecture of 3D convolution layer followed by ResNet-18 \cite{he2016resnet} for the visual front-end and two 2D convolution layers followed by one ResBlock for the audio front-end, following \cite{kim2021cromm}. Moreover, we use the encoder-decoder structure of \cite{ma2021end} that utilizes Conformer \cite{gulati2020conformer} for the encoder and Transformer \cite{serdyuk2022transformer} for the decoder, as shown in Fig.\ref{fig:1}. 

Firstly, the enhanced audio features from the V-CAFE is fused with the visual features which can be written as follows,
\begin{align}
    f_f = (\hat{f}_a \parallel f_v)W_f + b_f,
\end{align}
where $\parallel$ representss a concatenation, and $W_f\in\mathbb{R}^{2D_1\times D_1}$ and $b_f\in\mathbb{R}^{D_1}$ represents the weight and bias parameters for audio-visual fusion.
Then, the fused feature $f_f$ is succeeded to the input for the sequence encoder-decoder.
The whole network is trained with joint CTC/Attention loss function \cite{kim2017joint}. CTC \cite{graves2006connectionist} loss has a form of $p_{c}(y|x)\approx \Pi_{t=1}^Tp(y_t|x)$ with independence assumption between each output predictions and attention-based loss has a form of $p_{a}(y|x)=\Pi_{l=1}^Lp(y_l|y_{< l},x)$ that the current prediction is conditioned on both previous predictions and inputs. Therefore, the total loss function of joint CTC/Attention loss can be written as follows,
\begin{align}
    \mathcal{L} = w\log p_a(y|x) + (1-w) \log p_c(y|x),
\end{align}
where $w$ represents the balancing weight for the loss functions.

%------------------------------------ Table 1
%#####################################################
\input{./table/table1.tex}

\section{Experimental Setup}
\subsection{Datasets}
We utilize two large-scale audio-visual datasets, LRS2 and LRS3, to evaluate the effectiveness of the proposed method. For the audio stream in both datasets (16kHz), we use a hop size of 160, a window size of 400, and 80 mel-filter banks, so that the resulted log mel-spectrogram is 100 fps.
\vspace{-0.15cm}
\subsubsection{LRS2}
LRS2 dataset \cite{chung2017lrs2} is collected from BBC television shows. It consists of about 140,000 videos which are total about 224 hours long. Its videos are 25 fps and have frame size of 160$\times$160. For the pre-processing, we crop the frames with a size of 80$\times$80 centered at the lip, resize them to 112$\times$112, and convert the RGB colors to grayscale.
\vspace{-0.15cm}
\subsubsection{LRS3}
LRS3 dataset \cite{afouras2018lrs3} consists of about 150,000 videos which are total about 439 hours long and collected from TED and TEDx programs. The videos are 25 fps with 224$\times$224 resolution. During pre-processing, they are cropped centered at the mouth with a size of 100$\times$100, resized into 112$\times$112, and converted to grayscale, similar to LRS2.

\subsection{Implementation Details}
The first two convolution layers of the audio front-end reduce the frame length of audio streams by four times with stride convolutions. Thus, the resulting audio features from the audio front-end is 25 fps which is the same as the visual stream. For the Conformer \cite{gulati2020conformer} sequence encoder, we use hidden dimensions of 512, feed forward dimensions of 2048, 12 layers, 8 attention heads, and convolution kernel size of 31, and for the Transformer \cite{vaswani2017attention} sequence decoder, hidden dimensions of 512, feed forward dimensions of 2048, 6 layers, and 8 attention heads are employed. For the tokenizer, we use sentencepiece \cite{kudo2018sentencepiece} trained on both LRS2 and LRS3 corpus, with a vocabulary size of 4,000. The visual front-end is intialized using a pre-trained model from LRW \cite{chung2016lrw}, and the whole network is trained using pre-train sets of both LRS2 and LRS3 following \cite{zhang2019convlipreading,kim2021cromm,afouras2018deep,kim2022distinguishing}. Finally, the model is trained on its own dataset, LRS2 and LRS3. For the data augmentation purpose, horizontal random flipping and random spatial and temporal erasing are performed for the video stream. For the audio stream, we use SpecAugment \cite{park2019specaugment}, and acoustic noise in diverse environments of DEMAND dataset \cite{thiemann2013demand} is applied with random SNR levels. We use an initial learning rate of 0.0001 with AdamW \cite{kingma2014adam,loshchilov2017adamw} optimizer and batch size of 32 for training. $D_1$ and $D_2$ are set to 512 and 256, respectively, and $w$ is set to 0.8. During decoding, an external language model \cite{mikolov2010rnnLM} composed of two layered LSTM trained on Librispeech \cite{panayotov2015librispeech}, LRS2 \cite{chung2017lrs2}, and LRS3 \cite{afouras2018lrs3} are utilized, as follows,
\begin{align}
    \hat{y}=\text{argmax}((1-\alpha)\log p_a(y|x) &+ \alpha \log p_c(y|x) \\ &+ \beta \log p_{lm}(y)), \notag
\end{align}
where $\alpha$ is CTC weight, $p_{lm}(\cdot)$ is the language model prediction, and $\beta$ is relative weight for the language model. For the experiments, $\alpha$=0.3 and $\beta$=0.1 are used for LRS2 dataset, and $\alpha$=0.25 and $\beta$=0.15 are used for LRS3 dataset. We set the architecture of \cite{ma2021end} as our \textit{baseline} and reproduce the results with the same experimental settings as ours.

%------------------------------------ Figure 3
%#####################################################
\input{./figure/figure3.tex}
%#####################################################

\section{Experimental Results}

\subsection{Effectiveness of V-CAFE in noise suppression}
We verify the effectiveness of the proposed method by adding the V-CAFE onto the baseline model \cite{ma2021end} (\ie, architecture without the V-CAFE in Fig.\ref{fig:1}) which directly fuses the visual and audio features using concatenation. In order to confirm the noise-robustness of the V-CAFE, the experiments are conducted within different noise levels (\ie, SNR) of -5, 0, 5, 10, and 15dB. For the additive noise audio, we use \textit{`babble'} noise of NOISEX-92 \cite{varga1993noisex92} database which are not utilized during training. Table \ref{table:1} shows the comparison results on LRS2 and LRS3 datasets. By adopting the V-CAFE, the performances are improved for all SNR levels. Especially, when the SNR is lower so that the noise becomes stronger, the performance gap between the proposed method and the baseline becomes larger on both datasets. When the SNR is -5dB, the proposed method improves the performances from the baseline by about 2.8\% WER on LRS2 and 3.3\% WER on LRS3. This shows that the proposed V-CAFE can effectively suppress the noisy audio representations by incorporating the visual context information for the noise mask generation.

Moreover, to explore the noise-robustness of the different speech recognition methods, audio-based Automatic Speech Recognition (ASR), Visual Speech Recognition (VSR), and the proposed AVSR, we compare the performances of each method under different SNR levels of two additive noises of NOISEX-92 \cite{varga1993noisex92}, \textit{`factory1'} and \textit{`buccaneer1'}. Fig.\ref{fig:3} indicates graphs of WER as a function of the noise level on both LRS2 (Fig.\ref{fig:3}(a)) and LRS3 (Fig.\ref{fig:3}(b)) datasets. Since VSR model does not affect by the acoustic noise, it shows constant WER results of 37.9\% and 43.3\% on LRS2 and LRS3, respectively. When the SNR is high so the noise level is low, both ASR and AVSR methods have similar performances. However, when the input audio is strongly corrupted to the noise, the performance of ASR is significantly degraded and even worse than the visual-only model (\ie, VSR). For instance of the noise \textit{`buccaneer1'}, the performance gaps between ASR and AVSR are 39.30\%, 15.67\%, 5.37\%, 2.14\%, and 1.02\% for noise level -5, 0, 5, 10, and 15dB SNR, respectively, on LRS2, and 38.36\%, 13.35\%, 4.11\%, 1.43\%, and 0.35\% for noise level -5, 0, 5, 10, and 15dB SNR, respectively, on LRS3. The results verify that the AVSR model with the V-CAFE achieves the smallest performance degradation under the strong noise levels (\ie, -5dB and 0dB SNR) regardless of the noise sources. Therefore, to achieve noise-robust speech recognition, utilizing both audio and visual modalities is important, and considering the visual context can further boost the noise-robustness of the AVSR model.

%------------------------------------ Table 2
%#####################################################
\input{./table/table2.tex}
%#####################################################
%------------------------------------ Table 4
%#####################################################
\input{./table/table4.tex}
%#####################################################

\subsection{Comparison with speech enhancement methods}
For verifying the effectiveness of the proposed V-CAFE with diverse approaches of AVSR, we compare the V-CAFE with the previous speech enhancement approaches \cite{xu2020discriminative, hegde2021visual}. 
\textit{Pseudo Visual} \cite{hegde2021visual} is a method that enhances the noisy corrupted audio using pseudo visual information generated from the model. Then, the enhanced audio is recognized by a pre-trained ASR model. \textit{Discriminative} \cite{xu2020discriminative} is a two-stage AVSR approach that firstly enhances the audio spectrogram using audio-visual correspondence and then performs AVSR.
Table \ref{table:2} shows the WER comparisons in various noise levels on LRS2 and LRS3 databases. For the noise, we use \textit{`babble'} noise of NOIEX-92 \cite{varga1993noisex92}. The \textit{Pseudo Visual} method does not perform well under strong noise environment. This shows the limitation of the audio-only model (\ie, ASR) that even if the noise corrupted audio is enhanced by using visual information, recognizing using audio modality only cannot perform well in the strong noise situation. On the other hand, the \textit{Discriminative} and the proposed V-CAFE show robust performances to the acoustic noise by using audio and visual modalities simultaneously in speech recognition. Especially, the V-CAFE outperforms the previous method, \textit{Discriminative}, by about 3.6\% WER and 2.1\% WER at -5dB SNR on LRS2 and LRS3, respectively. The results confirm that the proposed architecture trained in an end-to-end manner with the visual context better performs in noisy conditions compared to the two-stage architectures, \textit{Pseudo Visual} \cite{hegde2021visual} and \textit{Discriminative} \cite{xu2020discriminative}.

\subsection{Results comparison with state-of-the-art methods}
We also report the performance of the proposed method in noise-clean setting compared with the state-of-the-art methods \cite{afouras2018deep, petridis2018avsrhybrid, yu2020overlaplrs2, xu2020discriminative, makino2019RNNt, ma2021end}, shown in Table \ref{table:4}. The results show that the proposed method is also effective in the noise-clean environment by achieving 4.3\% WER and 2.9\% WER on LRS2 and LRS3 datasets, respectively.

\subsection{Experiments on overlapped speech}
Finally, we verify the effectiveness of the proposed method in recognizing overlapped speech by training the AVSR models with the overlapped speech. We construct the overlapped speech by randomly selecting a sample in the same split but different from the input audio and overlapping those samples. Therefore, two different speeches are simultaneously played while the input video includes the target speech only. The proposed V-CAFE model achieves 20.00\%, 13.33\%, 9.94\%, 7.50\%, and 6.48\% on noise level -5, 0, 5, 10, and 15dB SNR, respectively on LRS2 dataset, as indicated in Table \ref{table:3}, attaining the highest performances over the \textit{baseline}. Moreover, consistent results for the LRS3 dataset can be found in the table. The experimental results confirm that the proposed AVSR method with the V-CAFE is not only effective for the environmental noise but also for the overlapped speech.

%------------------------------------ Table 3
%#####################################################
\input{./table/table3.tex}
%#####################################################
\vspace{-0.1cm}
\section{Conclusion}
In this paper, we design Visual Context-driven Audio Feature Enhancement module (V-CAFE) which enhances the noisy input speech with the aid of visual contexts obatined from the lip video. The proposed V-CAFE explicitly incorporates the visual context through cross-modal attention to generate a noise reduction mask, and the enhanced audio features are obtained from the mask. We conduct the experiments on various settings using LRS2 and LRS3 datasets in order to verify the effectiveness of the V-CAFE in noisy environments and show that it can further improve the noise-robustness of an AVSR model.
\vspace{-0.2cm}
\section{Acknowledgements}
This work was partially supported by the National Research Foundation of Korea (NRF) grant funded by the Korea government (MSIT) (No. NRF-2022R1A2C2005529) and Genesis Lab under a research project (G01210424).

% \clearpage
\bibliographystyle{IEEEtran}

\bibliography{mybib}

\end{document}

%% file: figure/figure1.tex
%------------------------------------ Figure 1
%#################################################
\begin{figure*}[t]
	\begin{minipage}[b]{1.0\linewidth}
		\centering
		\centerline{\includegraphics[width=15.0cm]{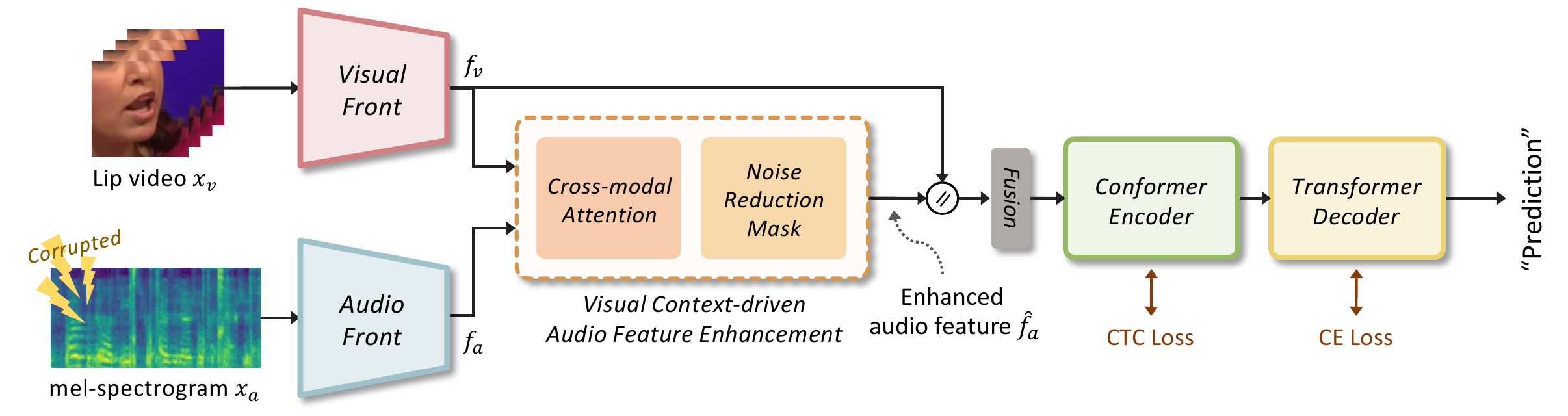}}
	\end{minipage}
	\caption{Overall architecture of the proposed method, containing Visual Context-driven Audio Feature Enhancement (V-CAFE).}
	\label{fig:1}
	\vspace{-0.4cm}
\end{figure*}
%##################################################

%% file: figure/figure2.tex
%------------------------------------ Figure 2
%#################################################
\begin{figure}[t]
	\begin{minipage}[b]{1.0\linewidth}
		\centering
		\centerline{\includegraphics[width=5.3cm]{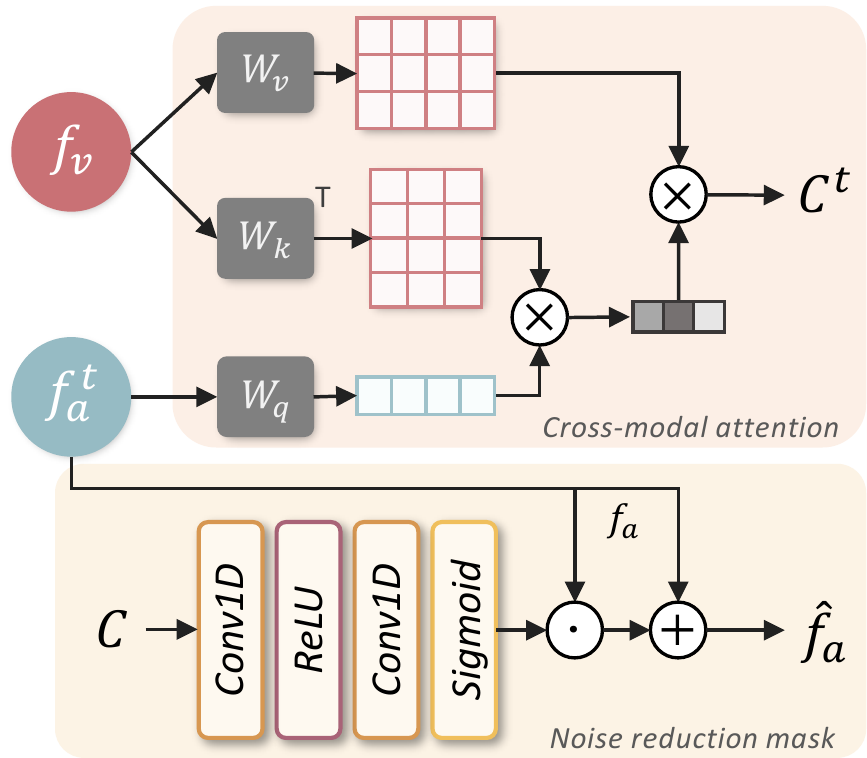}}
	\end{minipage}
	\caption{Visual Context-driven Audio Feature Enhancement.}
	\label{fig:2}
	\vspace{-0.5cm}
\end{figure}
%##################################################

%% file: table/table1.tex
\begin{table}[]
% \fontsize{11}{9}
	\renewcommand{\arraystretch}{1.2}
	\renewcommand{\tabcolsep}{2.5mm}
\caption{WER (\%) of baseline and the proposed V-CAFE on LRS2 and LRS3 datasets in different noise levels.}
\vspace{-0.1cm}
\centering
\resizebox{0.95\linewidth}{!}{%\large
\begin{tabular}{ccccc}
\Xhline{3\arrayrulewidth}
\multirow{2.5}{*}{\textbf{SNR (dB)}} & \multicolumn{2}{c}{\textbf{LRS2}}                                                                                                        & \multicolumn{2}{c}{\textbf{LRS3}}                                                                                                        \\ \cmidrule(l){2-3} \cmidrule(l){4-5}
                        & Baseline \cite{ma2021end} & \textbf{V-CAFE} & Baseline \cite{ma2021end} & \textbf{V-CAFE} \\  \cmidrule(l){1-3} \cmidrule(l){4-5}
-5 & 25.22 & \textbf{22.43} & 28.47   & \textbf{25.15} \\
0 & 12.44 &  \textbf{11.02} & 11.12 & \textbf{10.88} \\
5 & 6.94 &  \textbf{6.40}  & 5.94 & \textbf{5.73} \\
10 & 5.52 & \textbf{5.52}  & 4.33   & \textbf{4.09} \\
15  & 4.92 & \textbf{4.69} & 3.60 & \textbf{3.37} \\
clean & 4.63 & \textbf{4.33} & 3.16  & \textbf{2.94} \\ \Xhline{3\arrayrulewidth}
\vspace{-1.1cm}
\end{tabular}}
	\label{table:1}
\end{table}

%% file: figure/figure3.tex
%------------------------------------ Figure 1
%#################################################
\begin{figure}[t]
	\begin{minipage}[b]{1.0\linewidth}
		\centering
		\centerline{\includegraphics[width=8cm]{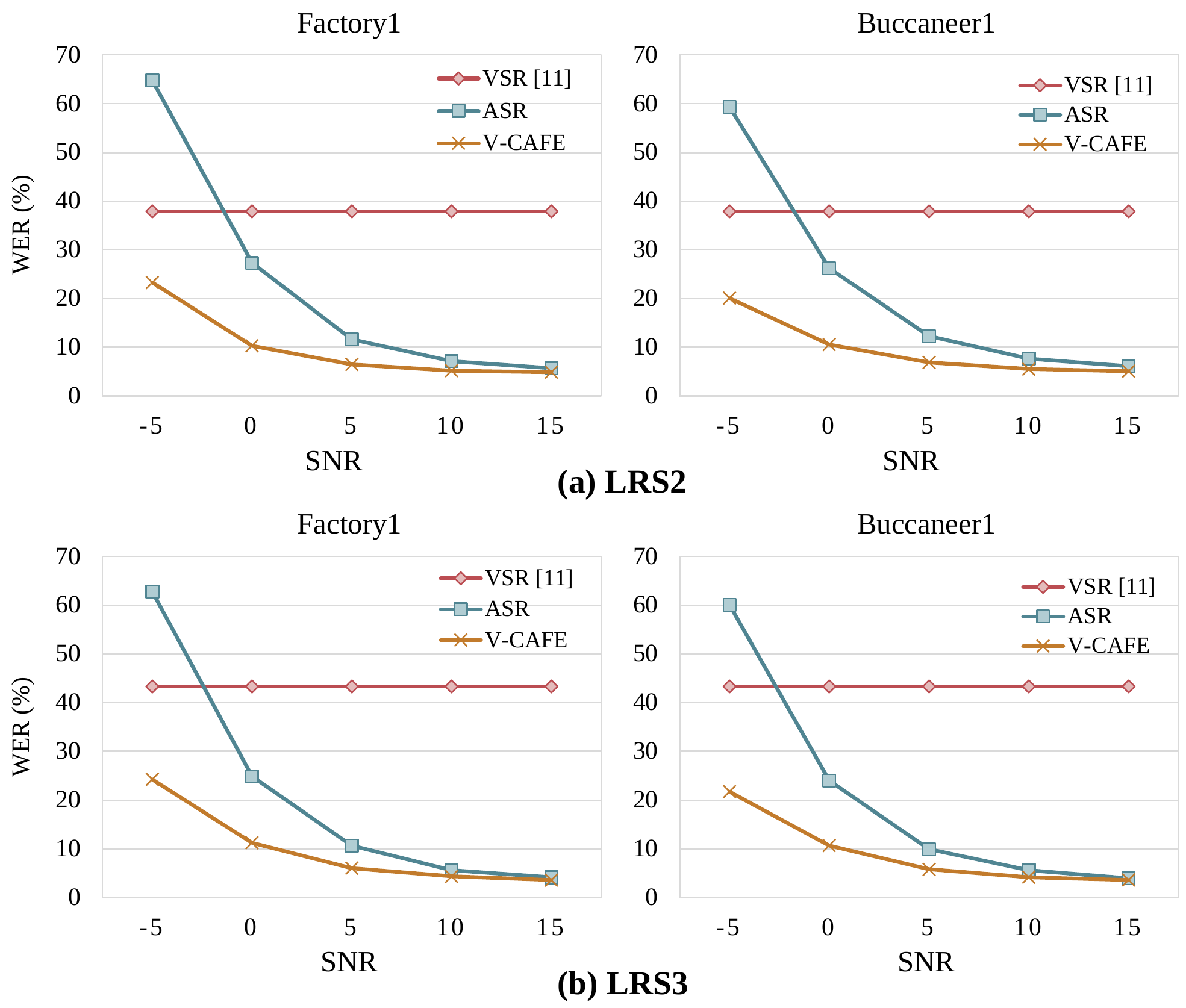}}
	\end{minipage}
	\vspace{-0.6cm}
	\caption{WER (\%) on two additive noises in different noise levels of (a) LRS2 and (b) LRS3 datasets.}
	\label{fig:3}
	\vspace{-0.5cm}
\end{figure}
%##################################################

%% file: table/table2.tex
% \begin{table}[]
% 	\renewcommand{\arraystretch}{1.3}
% 	\renewcommand{\tabcolsep}{2.3mm}
% \caption{WER performance comparisons with 2-stage architectures}
% \centering
% \vspace{-0.1cm}
% \resizebox{0.88\linewidth}{!}{
% \begin{tabular}{cccccc}
% \Xhline{3\arrayrulewidth}
% \textbf{Methodology} & -5 & 0 & 5 & 10 & clean \\ \hline
% Discrimitive \cite{xu2020discriminative} & &  &  & \\
% Visualvoice \cite{gao2021visualvoice} & & &  &  \\ \hdashline
% \textbf{V-CAFE} & & & & \\ %\hdashline
% \Xhline{3\arrayrulewidth}
% \end{tabular}}
% % 	\vspace{-0.1cm}
% 	\vspace{-0.4cm}
% 	\label{table:6}
% \end{table}

\begin{table}[]
	\renewcommand{\arraystretch}{1.2}
	\renewcommand{\tabcolsep}{2.3mm}
\caption{WER (\%) comparisons with speech enhancement architectures.}
\centering
\vspace{-0.2cm}
\resizebox{0.9999\linewidth}{!}{%\normalsize
\begin{tabular}{ccccccc}
\Xhline{3\arrayrulewidth}
\multirow{2.5}{*}{\textbf{Dataset}} & \multirow{2.5}{*}{\textbf{Method}} & \multicolumn{5}{c}{\textbf{SNR (dB)}} \\ \cmidrule(l){3-7}
                         &                              & -5  & 0  & 5  & 10 & 15 \\ \midrule
\multirow{3}{*}{\textbf{LRS2}}  & Pseudo Visual \cite{hegde2021visual} & 78.73 & 39.75 & 17.08 & 9.30 & 7.42      \\ 
& Discriminative \cite{xu2020discriminative}     & 26.09& 12.59 & 7.05 & \textbf{5.29} & 4.86      \\ 
                         %& Visualvoice \cite{gao2021visualvoice}                 &  &  &  &  &       \\ 
                         
                         & \textbf{V-CAFE}              & \textbf{22.43} & \textbf{11.02} &\textbf{ 6.40} &5.52& \textbf{4.69}      \\ \midrule
\multirow{3}{*}{\textbf{LRS3}}  & Pseudo Visual \cite{hegde2021visual}  & 78.28 & 35.63 & 12.46 & 6.64 & 4.43      \\%\cmidrule(l){2-7}
& Discriminative \cite{xu2020discriminative}     &  27.27 & 12.16  & 5.83 & \textbf{3.92} & \textbf{3.28}      \\
                         %& Visualvoice \cite{gao2021visualvoice}                 &     &    &    &    &       \\
                         
                         & \textbf{V-CAFE }             & \textbf{25.15}&\textbf{10.88}& \textbf{5.73}  & 4.09  &    3.37  \\ 
                         \Xhline{3\arrayrulewidth}
\end{tabular}}
	\vspace{-0.3cm}
	\label{table:2}
\end{table}

%% file: table/table4.tex
\begin{table}[]
	\renewcommand{\arraystretch}{1.3}
	\renewcommand{\tabcolsep}{5.7mm}
\centering
\caption{WER (\%) comparisons with state-of-the-art methods.}
\vspace{-0.2cm}
\resizebox{0.76\linewidth}{!}{\normalsize
\begin{tabular}{ccc}
\Xhline{3\arrayrulewidth}
\textbf{Method}  & \textbf{LRS2} & \textbf{LRS3} \\ \hline
TM-Seq2Seq \cite{afouras2018deep}  & 8.5 & 7.2 \\ 
CTC/Attention \cite{petridis2018avsrhybrid} & 7.0 & - \\ 
LF-MMI TDNN \cite{yu2020overlaplrs2} & 5.9 & -  \\ 
EG-Seq2Seq \cite{xu2020discriminative}  & - & 6.8 \\
RNN-T \cite{makino2019RNNt} & - & 4.5 \\ \hdashline
% Conformer\footnotemark \cite{ma2021end}& 4.6 & 3.1 \\\hline
Baseline \cite{ma2021end}& 4.6 & 3.2 \\\hline
% Conformer \cite{ma2021end} & 3.7 & 2.3 \\ \hline
\textbf{V-CAFE}  & \textbf{4.3}& \textbf{2.9} \\
\Xhline{3\arrayrulewidth}
\vspace{-1.1cm}
\end{tabular}}
	\label{table:4}
\end{table}
%\footnotetext{Reproduced results with the same front-end and language model with the proposed method.}

%% file: table/table3.tex
\begin{table}[]
	\renewcommand{\arraystretch}{1.3}
	\renewcommand{\tabcolsep}{2.8mm}
\caption{WER (\%) comparisons on overlapped speech.}
\centering
\vspace{-0.2cm}
\resizebox{0.9999\linewidth}{!}{%\normalsize
\begin{tabular}{ccccccc}
\Xhline{3\arrayrulewidth}
\multirow{2.5}{*}{\textbf{Dataset}} & \multirow{2.5}{*}{\textbf{Method}} & \multicolumn{5}{c}{\textbf{SNR (dB)}} \\ \cmidrule(l){3-7}
                         &                              & -5  & 0  & 5  & 10 & 15 \\ \hline
\multirow{2}{*}{\textbf{LRS2}}  & Baseline \cite{ma2021end} & 24.90 & 16.51 & 10.83 & 7.89 & 6.54      \\ 
                         & \textbf{V-CAFE}              & \textbf{20.00} & \textbf{13.33} &\textbf{9.94} & \textbf{7.50} & \textbf{6.48}      \\ \hline
\multirow{2}{*}{\textbf{LRS3}}  & Baseline \cite{ma2021end}  & 22.35 & 13.80 & 9.43 & 6.47 & 4.97      \\

                         & \textbf{V-CAFE }             & \textbf{19.30}&\textbf{12.50}& \textbf{8.34}  & \textbf{5.42}  &   \textbf{4.41}  \\ 
                         \Xhline{3\arrayrulewidth}
\end{tabular}}
\vspace{-0.6cm}

	\label{table:3}
\end{table}